\documentclass[12pt]{iopart}
\usepackage{citesort}
\usepackage{braket}
\usepackage{graphicx}
\usepackage{textgreek}
\newcommand{\abs}[1]{\left|#1\right|}
\newcommand{\ketbra}[1]{{|{#1}\rangle}\!{\langle{#1}|}}
\renewcommand{\Re}{\textnormal{Re}}
\renewcommand{\Im}{\textnormal{Im}}

\begin{document}
\title[]{Precise tomography of optical polarization qubits under conditions of chromatic aberration of quantum transformations}

\author{B.I.~Bantysh, Yu.I.~Bogdanov, N.A.~Bogdanova, Yu.A.~Kuznetsov}

\address{Valiev Institute of Physics and Technology of Russian Academy of Sciences, Moscow, Russia}
\ead{bbantysh60000@gmail.com}
\vspace{10pt}
\begin{indented}
\item[]January 2020
\end{indented}

\begin{abstract}
In this work we present an algorithm of building an adequate model of polarizing quantum state measurement. This model takes into account chromatic aberration of the basis change transformation caused by the parasitic dispersion of the wave plates crystal and finite radiation spectral bandwidth. We show that the chromatic aberration reduces the amount of information in the measurements results. Using the information matrix approach we estimate the impact of this effect on the qubit state reconstruction fidelity for different values of sample size and spectral bandwidth. We also demonstrate that our model outperforms the standard model of projective measurements as it could suppress systematic errors of quantum tomography even when one performs the measurements using wave plates of high order.
\end{abstract}
\maketitle

\section{Introduction}

The light polarization has a wide range of applications for the tasks of classical optics (polarizing filters, liquid--crystal display, phase modulators) and quantum information technologies (quantum networks, quantum memory, quantum computations and quantum simulators) \cite{goldstein2017,collet2005,slussarenko2019,liao2017}. The mathematical apparatus of quantum polarizing optics is based on quantum mechanical description of vector states in a Hilbert space. In general, the measurement of such states could be described in terms of POVM-measurements (positive--operator valued measure) \cite{nielsen2000}.

Practically the polarization measurement in the computational basis ($\ket{0}=\ket{V}$, $\ket{1}=\ket{H}$) could be performed by the light transmission through the polarizing beam splitter and the light detection at both outputs. However, such a measurement is not enough to get access to the complex amplitudes of an arbitrary state of superposition $\ket{\psi}=c_0\ket{0}+c_1\ket{1}$. According to N.~Bohr's complementarity principle \cite{bohr1996} the unknown quantum state reconstruction implies performing a set of complementary measurements in different bases. \textit{Quantum tomography} procedure is reconstruction of a quantum state using the results of mutually complementary measurements \cite{lvovsky2009,paris2004,dariano2003,banaszek2013,bogdanov2009,bogdanov2010,bogdanov2011}.

\begin{figure}[h]
  \centering
  \includegraphics[width=.4\linewidth]{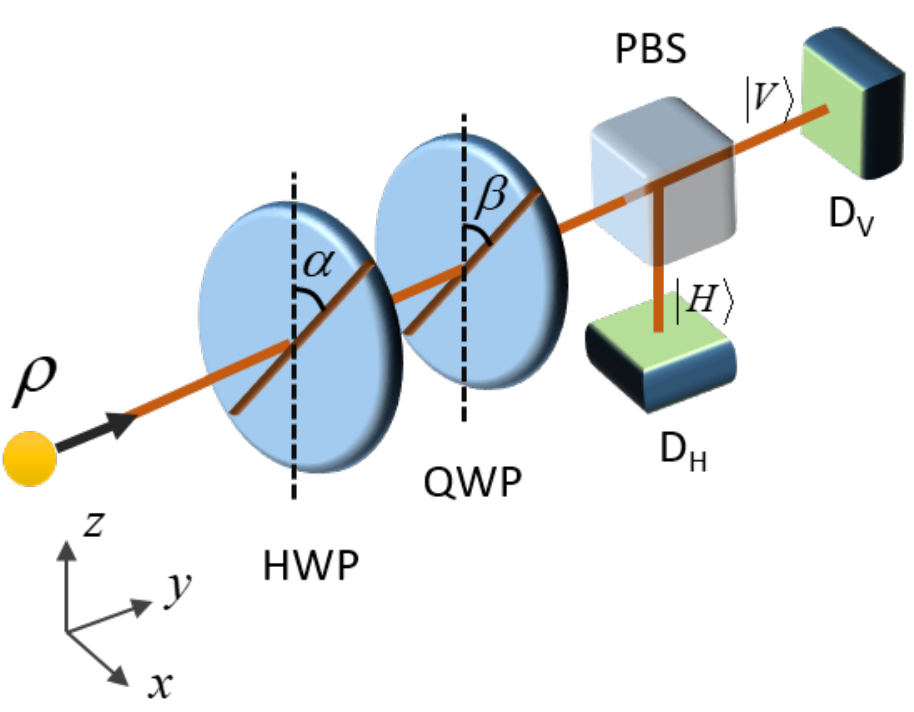}
  \caption{Polarizing state measurement in an arbitrary basis: HWP --- half--wave plate, QWP --- quarter--wave plate, PBS --- polarizing beam splitter, D\textsubscript{V} and D\textsubscript{H} --- photo detectors.}
  \label{fig:scheme}
\end{figure}

To change the polarization measurement basis one may use a set of two wave plates: a half-wave plate (HWP) and a quarter-wave plate (QWP) oriented by angles $\alpha$ and $\beta$ to the vertical axis (\Fref{fig:scheme}). Wave plate is made of birefringent crystal. Its thickness is selected according to the radiation wave length. In terms of quantum information the polarizing state transformation is equivalent to the qubit rotation on Bloch sphere. However, the presence of radiation spectral degree of freedom results in chromatic aberration of qubit rotation because the transformation differs for each spectral component \cite{ghosh1999,bogdanov2013,bogdanov2014}. Such effects could lead to systematic errors and the reduction of quantum tomography accuracy \cite{williams1999,hou2016}. Combating the phenomenon is particularly important for manipulating polarizing cluster states \cite{vallone2008}.

Based on fuzzy measurements approach \cite{bogdanov2016,bogdanov2018,bantysh2019} we develop a method for the polarizing quantum states reconstruction that is robust to the chromatic aberration of the basis change transformation. These results could help eliminate systematic errors of quantum tomography caused by this effect.

The paper is organized as follows. In \Sref{sect:measurements} we consider complementary measurements and describe the origination of systematic errors due to probabilistic uncertainty of basis change transformation. In \Sref{sect:chrom} we present a simple algorithm for the construction of measurement operators in the case of light chromatic aberration inside the wave plates. In \Sref{sect:results} we perform theoretical analysis of completeness and fidelity of quantum measurements protocols for different values of spectral bandwidth. We perform a quantitative description showing the loss of information caused by chromatic aberration. We also use Monte Carlo simulation to compare our method to a standard one based on the model of ideal projective measurements. We show that this model does not give statistically adequate results and is limited in accuracy.

\section{Complementary measurements}\label{sect:measurements}

Let us consider a vector state $\ket{\psi}$ in a Hilbert space of dimension $s=2$ (the following could be naturally generalized to larger dimensions). The measurement in the basis $\{\ket{0},\ket{1}\}$ brings us to the consideration of a binomial distribution (``quantum coin'') with probabilities $p_0=\abs{\braket{0|\psi}}^2$ and $p_1=\abs{\braket{1|\psi}}^2$. Obviously this distribution is not enough to measure (reconstruct) complex amplitudes of the state $\ket{\psi}$. According to the N.~Bohr's complementarity principle one must obtain a set of mutually complementary distributions. To get one of these distributions let us transform the initial state $\ket{\psi}$ by a unitary transformation $U$ and then perform a measurement. The resulting binomial distribution with probabilities $\tilde{p}_0=\abs{\braket{0|U|\psi}}^2 = \abs{\braket{\tilde{0}|\psi}}^2$ and $\tilde{p}_1=\abs{\braket{1|U|\psi}}^2 = \abs{\braket{\tilde{1}|\psi}}^2$ corresponds to the measurement of the state $\ket{\psi}$ in the new basis $\{\ket{\tilde{0}} = U^\dagger\ket{0}, \ket{\tilde{1}} = U^\dagger\ket{1}\}$. More general, the measurement of an arbitrary mixed state $\rho$ gives the probability distribution:
\begin{equation}\label{eq:prob}
  \tilde{p}_j = \Tr(U \rho U^\dagger P_j) = \Tr(\rho \tilde{P}_j), \quad j = 0,1
\end{equation}
with projectors $P_j=\ketbra{j}$, $\tilde{P}_j=U^\dagger P_j U = \ketbra{\tilde{j}}$.

The most relevant types of mutually complementary probability distributions are the ones that provide informational completeness. The measurement protocols that provide such distributions are sufficient for the reconstruction of an arbitrary quantum state \cite{bogdanov2010}. However, it could be practically difficult to perform an ideal ``sharp'' unitary transformation $U$. Any realization of $U$ would be more or less uncertain. In the case of an optical polarizing state ($\ket{0}$ and $\ket{1}$ correspond to the states of vertical and horizontal polarization respectively) $U$ could be implemented using a set of wave plates made of birefringent crystal. Due to chromatic dispersion of the crystal every spectral component of the radiation is subject to a slightly different transformation. This creates an uncertain (\textit{fuzzy}) picture of $U$. Note that it is a direct analogue of a classical chromatic aberration due to the parasitic light dispersion resulting in image defocusing.

Below we consider the particular type of chromatic aberration caused by parasitic dispersion of wave plates transformation and the finite radiation spectral bandwidth. At the same time, our method is general and could be adapted to account fuzzy transformation of any given nature.

\section{Measurement operators in the case of chromatic aberrations}\label{sect:chrom}

A half--wave (HWP) and a quarter--wave plates (QWP) are enough to measure polarizing state in an arbitrary basis (\Fref{fig:scheme}). In the ideal case of monochromatic light the plates perform the transformation $U(\alpha,\beta) = U_{WP}(\delta_{QWP},\beta) \cdot U_{WP}(\delta_{HWP},\alpha)$ where
\begin{equation}\label{eq:WP}
  U_{WP}(\delta,\alpha) = \left(\begin{array}{cc}
    \cos\delta-i\sin\delta\cos2\alpha & -i\sin\delta\sin2\alpha \\
    -i\sin\delta\sin2\alpha & \cos\delta+i\sin\delta\cos2\alpha
  \end{array}\right)
\end{equation}
is a wave plate Jones matrix, $\alpha$ --- angle between vertical axis and wave--plate fast axis, $\delta = \pi h |n_o-n_e|/\lambda$ --- relative optical thickness, $h$ --- geometrical thickness, $n_o$ and $n_e$ --- ordinary and extraordinary rays refractive indices respectively. Wave plates thickness are made as such, so that at the central wave length one would have $\delta_{HWP} = \pi/2 + \pi k$, $\delta_{QWP} = \pi/4 + \pi k$ ($k$ --- wave plates order). By the appropriate choice of $\alpha$ and $\beta$ one could implement any complete measurements protocol (\ref{app:protocols}). Note that in terms of quantum information the transformation \eref{eq:WP} corresponds to the qubit rotation on the Bloch sphere by an angle $2\delta$ around the axis $\vec{n}=(\sin 2\alpha, 0, \cos 2\alpha)$.

Let $P(\lambda_k)d\lambda$ be the probability of photon to have the wave length within the small $d\lambda$ neighborhood of $\lambda_k$. As every wave length value corresponds to a different value of relative optical thickness $\delta_k$ one obtains a chromatic aberration of the transformation: $\rho \rightarrow \sum_{k}{U_k(\alpha,\beta) \rho U_k^\dagger(\alpha,\beta) P(\lambda_k)d\lambda}$. Using \eref{eq:prob} and the trace operator cyclic permutations property one could obtain the following measurement operators:
\begin{equation}\label{eq:fuzzy_ops}
  \Lambda_j(\alpha,\beta) = \sum_{k}{U_k^\dagger(\alpha,\beta) P_j U_k(\alpha,\beta) P(\lambda_k)d\lambda}, \quad j = 0,1.
\end{equation}
These operators describe a fuzzy measurement of the initial state $\rho$ instead of ideal projectors $\tilde{P}_j$ which do not take chromatic aberration into account. As projectors form the composition of unity $P_0+P_1=E$ ($E$ is an identity operator) fuzzy operators also satisfy the same relation: $\Lambda_0 + \Lambda_1 = E$. Operators \eref{eq:fuzzy_ops} have the form of ideal projectors $\tilde{P}_j$ in the limit of monochromatic light and become POVM--operators in the finite spectral bandwidth case.

\section{Simulation results}\label{sect:results}

Below we illustrate the results of the simulation of polarizing states quantum tomography. HWP and QWP have the 10-\textit{th} order at the central wave length $\lambda_0 = 650$nm. The quartz wave plates thickness are $h_{HWP} = 756$\textmu m, $h_{QWP} = 738$\textmu m (the refractive indices of quartz were taken from the work \cite{ghosh1999}). We compare quantum state reconstruction using two measurement models: the standard model with ideal projectors and the more realistic one with fuzzy operators \eref{eq:fuzzy_ops} calculated for the bandwidth $\Delta \lambda$ (we assume spectrum to be uniform). We consider the measurement protocol with the cube symmetry. This protocol consists of three wave plates configurations with corresponding angles $\{(5\pi/8,\pi/2),(11\pi/16,3\pi/4),(\pi/2,\pi/2)\}$ (\ref{app:protocols}). This protocol is equivalent to the independent measurements of three Pauli observables. The simulation results for the protocol with octahedron symmetry are presented in \ref{app:octahedron}.

The results below are based on the informational fidelity theory and root approach to quantum tomography \cite{bogdanov2009,bogdanov2010} (see \ref{app:root} and \ref{app:fidelity} for details). MATLAB realization of these methods is available at \cite{matlablib}.

We estimate the accuracy of a pure quantum state reconstruction using standard \textit{fidelity} between two vector states: $F = \abs{\braket{\psi | \varphi}}^2$.

\subsection{Distribution of loss function}

Let us consider the loss function $L = n_{tot}\left<1-F\right>$, where $\left<1-F\right>$ is the average infidelity over different tomography experiments and $n_{tot}$ is the total sample size over all the wave plates configurations in a single experiment. $L$ depends on the measurement protocol and the quantum state under research. Obviously lower values of $L$ correspond to higher fidelities.

\begin{figure}[h]
  \centering
  \includegraphics[width=.7\linewidth]{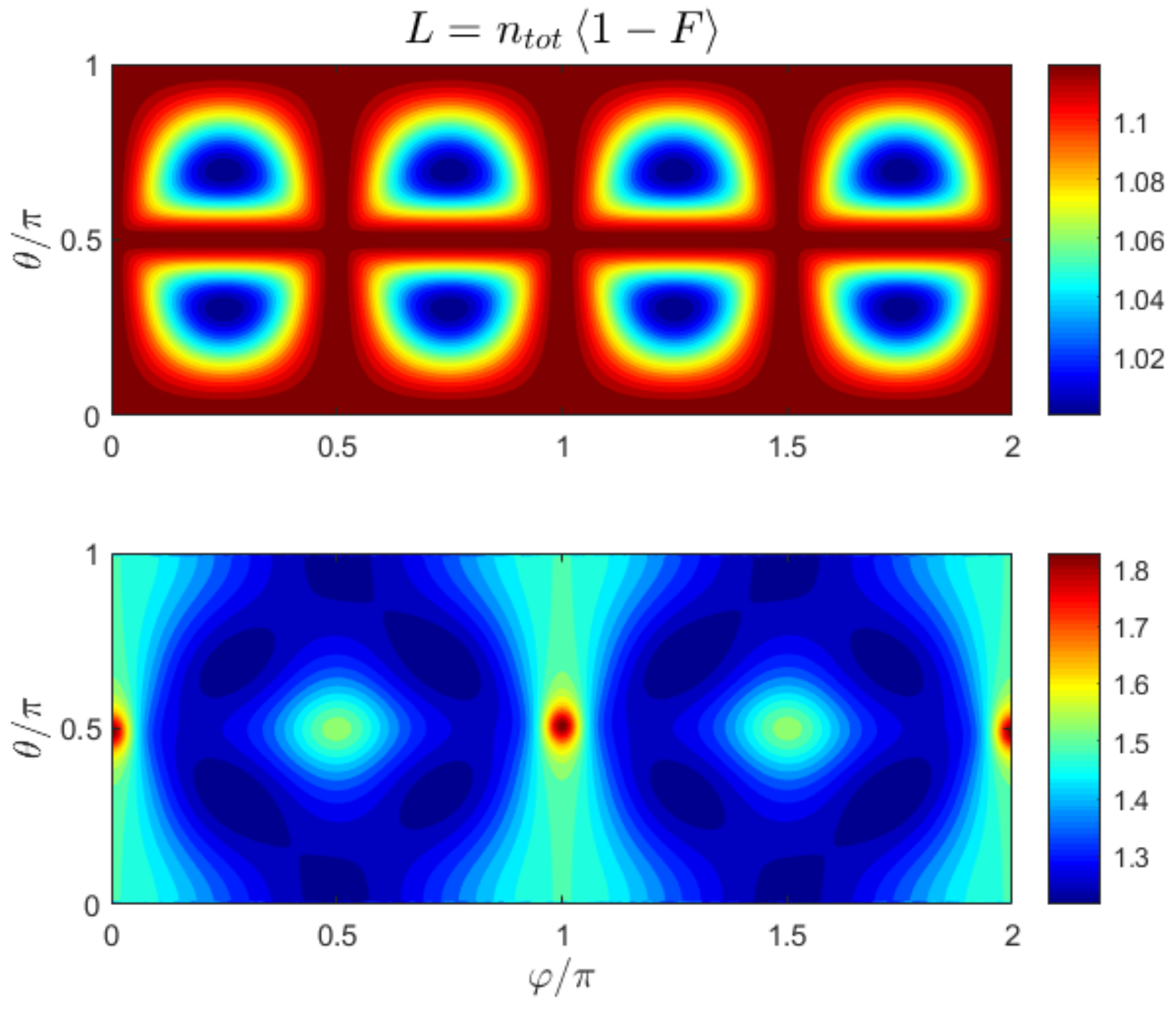}
  \caption{Loss function distribution over pure single-qubit states $\cos(\theta/2)\ket{0}+\sin(\theta/2)e^{i\varphi}\ket{1}$. Top --- the case of monochromatic radiation ($\Delta \lambda = 0$), bottom --- $\Delta \lambda = 0.01$\textmu m.}
  \label{fig:LBloch_cube}
\end{figure}

\Fref{fig:LBloch_cube} shows the loss function corresponding to the tomography of pure single-qubit states parameterized with spherical angles: $\ket{\psi} = \cos(\theta/2)\ket{0}+\sin(\theta/2)e^{i\varphi}\ket{1}$. For the ideal case of monochromatic light minimal and maximal values are $L_{min}=1$ and $L_{max}=1.125$ respectively \cite{bogdanov2011}. Note that $L_{min}=1$ is the theoretical minimum for any complete POVM--measurements of the single-qubit state.

The fuzzy nature of wave plates transformations caused by chromatic aberration results in the loss of information. The loss function becomes knowingly higher than the theoretical minimum equal to 1. For the case $\Delta\lambda = 0.01$\textmu m one has $L_{min} = 1.216$, $L_{max} = 1.857$. The distribution is also rearranged as a whole. Now the minimal fidelity loss ($L=1.216$) is even higher than maximal one for the case of monochromatic light ($L=1.125$). Let us also consider an example of the state $\sqrt{0.999}\ket{0}+i\sqrt{0.001}\ket{1}$. For the case of monochromatic light one would have $L\approx1.125$ corresponding to theoretical maximum, while in the presence of chromatic aberration effect one would obtain $L=1.235$ close to a minimum $L_{min}=1.216$.

\begin{figure}[h]
  \centering
  \begin{minipage}{0.48\textwidth}
    \centering
    \includegraphics[width=\textwidth]{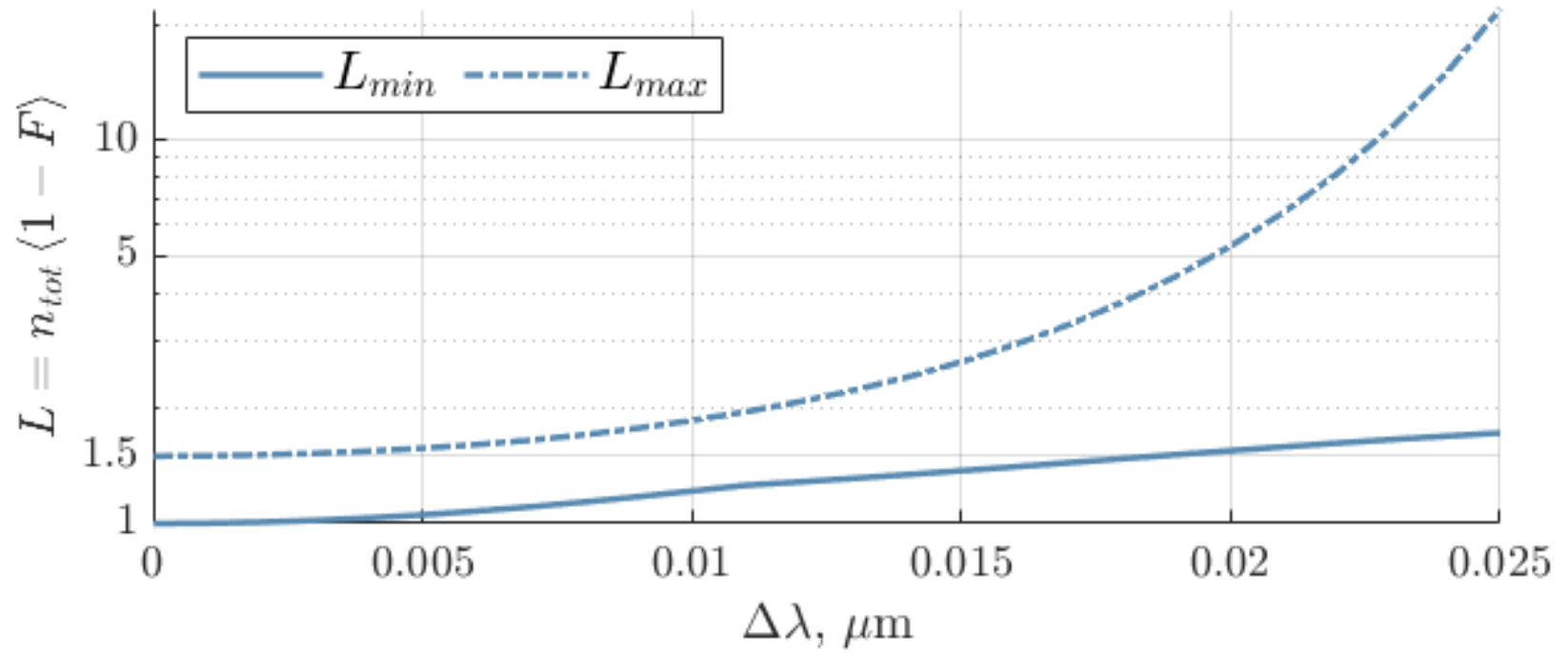}\\(a)
  \end{minipage}\hfill
  \begin{minipage}{0.48\textwidth}
    \centering
    \includegraphics[width=\textwidth]{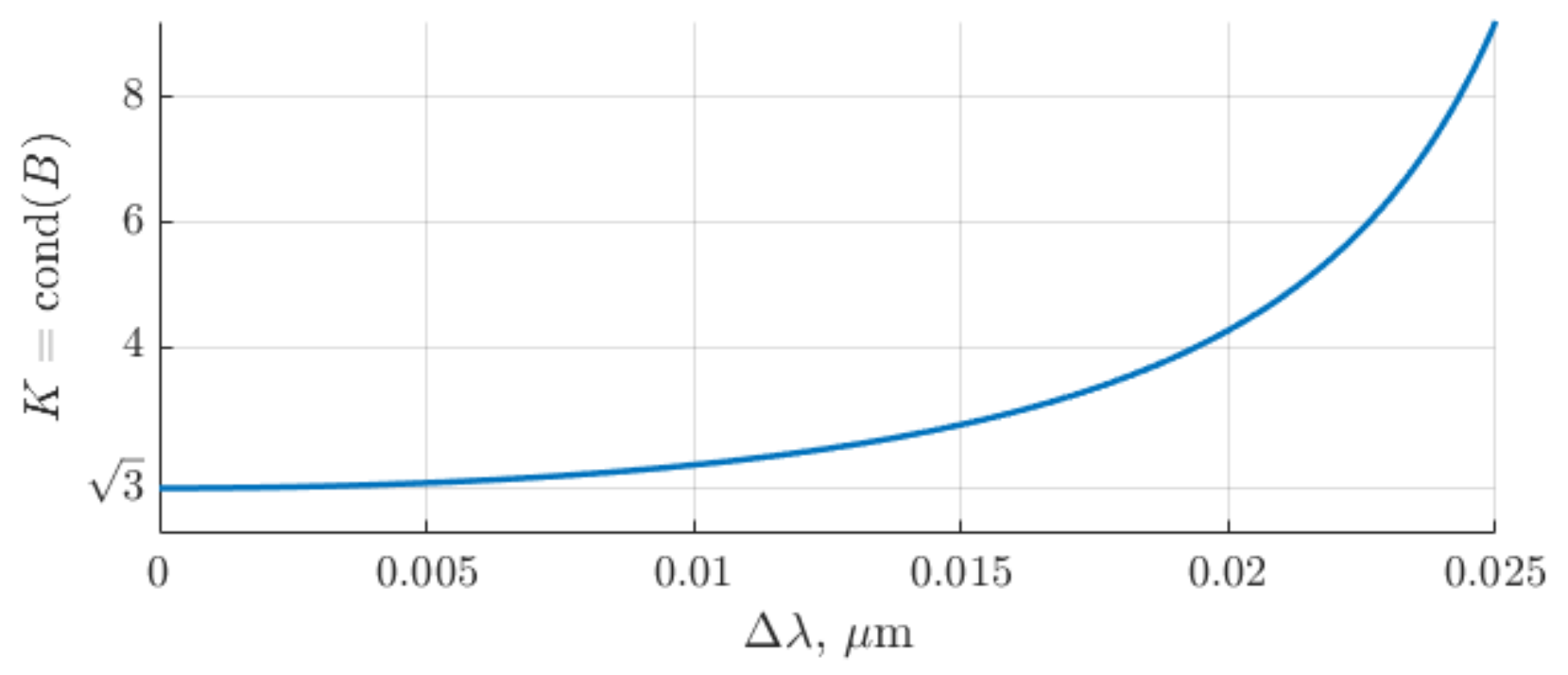}\\(b)
  \end{minipage}
  \caption{Minimal and maximal fidelity loss function (a) and measurement matrix condition number (b) vs. the radiation spectral bandwidth.}
  \label{fig:Lminmax_cond_cube}
\end{figure}

\Fref{fig:Lminmax_cond_cube}(a) shows the increase of minimal and maximal values of fidelity loss function with the increase of spectral bandwidth. Note that the value $L_{max}=1.5$ when $\Delta\lambda \rightarrow 0$ differs from the theoretical maximum 1.125 of ideal projective measurements ($\Delta\lambda = 0$). For $\Delta\lambda \rightarrow 0$ the deviation appears only within infinitesimal neighborhoods of eigenvectors of $\sigma_x$ and $\sigma_y$ Pauli operators.

\subsection{Measurement protocol completeness}

Let us convert the matrix of the measurement operator $\Lambda_j$ into a single row (the second matrix row is put to the right of the first one and so on). Let us perform this operation for all $s$ operators for each of $l$ wave plates configurations and join them into the measurement matrix $B$ of dimension $ls \times s^2$. The tomographic completeness of a measurement protocol means that matrix $B$ should have at least $s^2$ rows and contain no zero singular values \cite{bogdanov2010}. The condition number $K$ of matrix $B$ serves as the measure of protocol robustness towards statistical fluctuations. For large $K$ small fluctuations of experimental counts lead to the significant change of state reconstruction result and small fidelity.

Measurement matrix for the single-qubit cube protocol has dimension $6\times4$ ($s = 2$, $l = 3$). For a monochromatic light one would have $K=\sqrt{3}$. \Fref{fig:Lminmax_cond_cube}(b) shows the increase of condition number with the increase of the radiation spectral bandwidth.

\subsection{Numerical experiments}

\begin{figure}[h]
  \centering
  \begin{minipage}{0.48\textwidth}
    \centering
    \includegraphics[width=\textwidth]{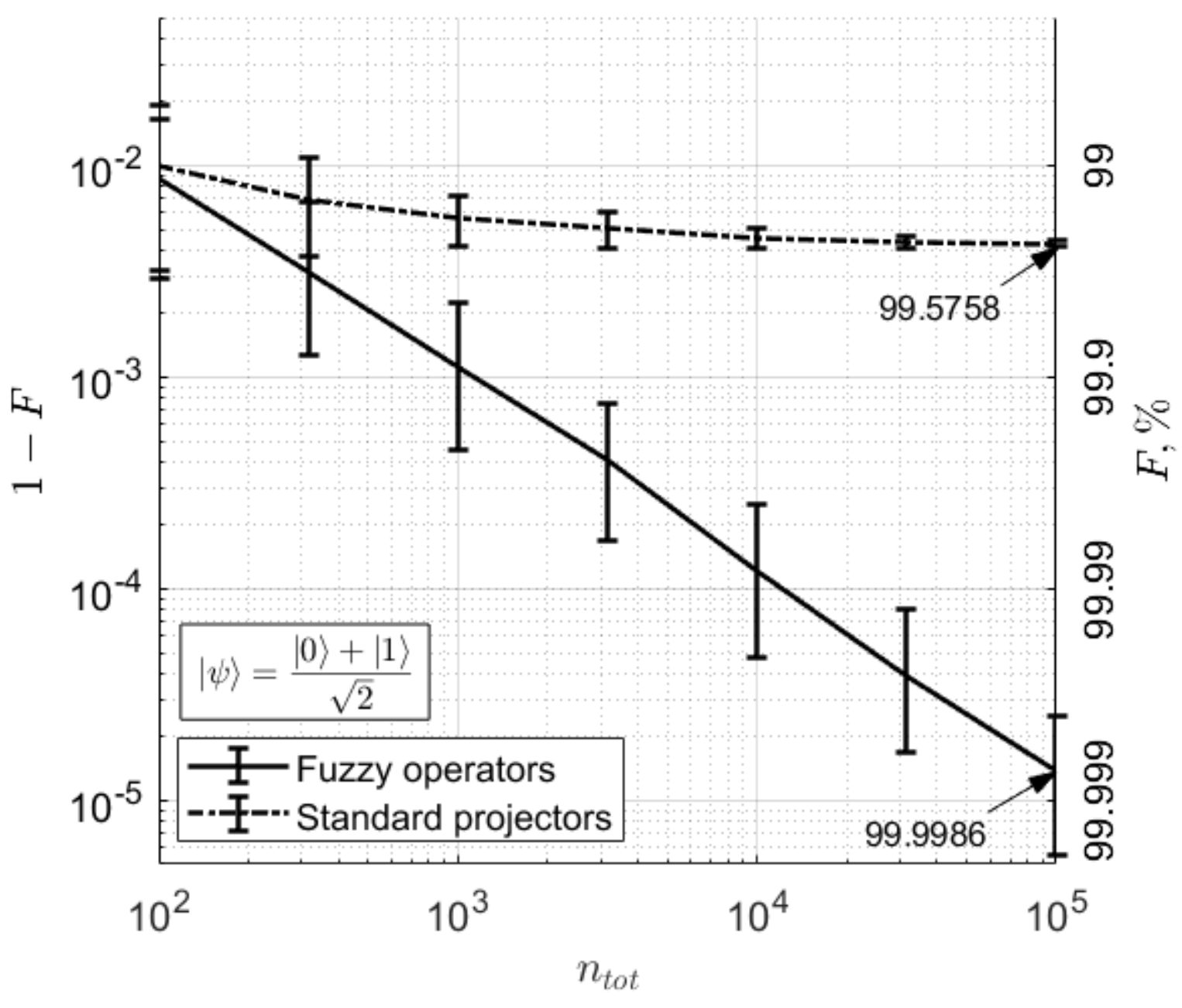}\\(a)
  \end{minipage}\hfill
  \begin{minipage}{0.48\textwidth}
    \centering
    \includegraphics[width=\textwidth]{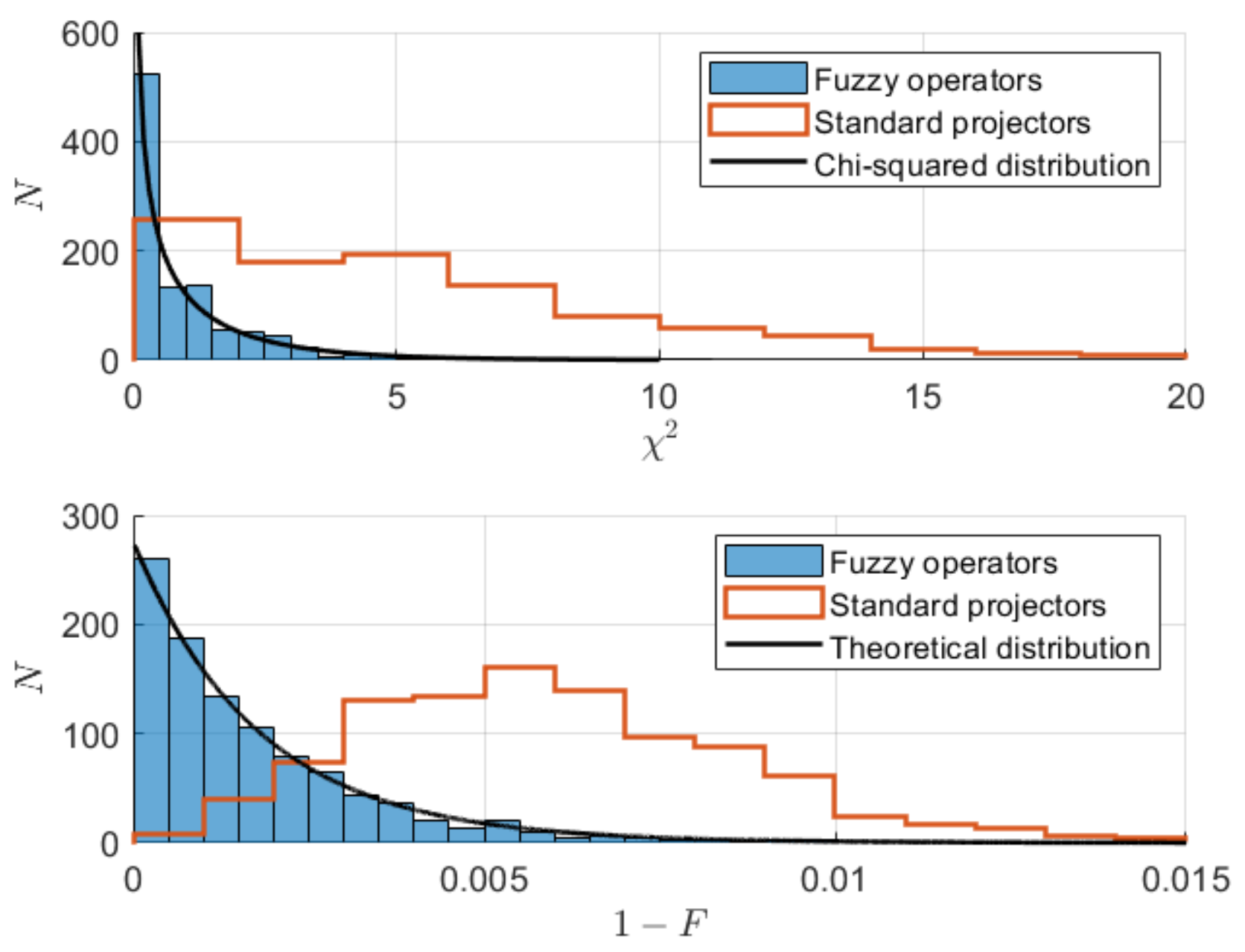}\\(b)
  \end{minipage}
  \caption{The results of Monte Carlo simulation of the quantum state tomography in presence of chromatic aberration. Radiation spectral bandwidth $\Delta\lambda = 0.01$\textmu m. Comparison of fuzzy measurements model at $\Delta\lambda = 0.01$\textmu m and standard model of projective measurements. (a) The reconstruction fidelity vs. total sample size. The plot shows median and lower and upper quartile over 1000 numerical experiments. (b) The result of 1000 numerical experiments with $n_{tot}=1000$ in each. Top --- histograms of chi-squared for two models and theoretical chi--squared distribution for 1 degree of freedom. Bottom --- histograms of infidelity for two models and theoretical distribution based on information matrix.}
  \label{fig:fid_cube}
\end{figure}

We performed the Monte Carlo simulation of quantum state complementary measurements. The results were then used to reconstruct quantum state using two measurement models: fuzzy measurements model with measurements operators \eref{eq:fuzzy_ops} and standard model of projective measurements. The results were obtained for the input state $\ket{\psi} = (\ket{0}+\ket{1})/\sqrt2$ --- one of those with the highest fidelity losses at $\Delta \lambda = 0.01$\textmu m (according to the results at Figure \Fref{fig:LBloch_cube}). We performed the reconstruction using root approach and maximum likelihood estimation (\ref{app:root}). The results demonstrate an adequacy of the fuzzy model and an inadequacy of the standard one (Figure \ref{fig:fid_cube}). It is important to note that for our fuzzy measurements model the observed values of infidelity are in a very close agreement with the theoretical limit obtained from the information matrix. The standard model gives much higher values of infidelity. In particular, one could see that the use of the standard model leads to the fidelity saturation at the level of about 99.58\%. This shows that the model is not quite adequate. At the same time, the use of fuzzy measurements operators could potentially give any desired level of fidelity by increasing the sample size.

Note that the quality of quantum state reconstruction with standard model greatly depends on radiation spectral bandwidth and wave plates order. The use of effective zero--order wave plates could significantly suppress systematic errors caused by chromatic aberration effect. In this case errors occur only for large enough sample size.

Also note that the fuzzy measurements operators formation implies having knowledge about the radiation spectral distribution and wave plates crystal dispersion relations. Inaccuracies in this knowledge could lead to some systematic errors originating for large samples.

\section{Conclusions}

Quantum state tomography implies performing a set of complementary measurements. To change the measurement bases one should transform the quantum state. However, in real experiments any realization of this transformation would be more or less uncertain. In particular, the tomography of polarizing quantum states implies the use of wave plates to change the basis. This transformation slightly differs for different spectral component --- a chromatic aberration of basis change transformation occurs. Such measurements could not be described in terms of projective measurements. Instead of projectors we construct POVM--measurements operators that take the considered effect into account. These operators could potentially provide an arbitrary high fidelity even for the wave plates of a high order.

Using the universal fidelity theory we analyzed the impact of chromatic aberration impact on the quantum tomography accuracy characteristics. We observed a significant change of loss function distribution with the increase of spectral bandwidth. We also showed that the use of standard model of projective measurements results in statistically inadequate results and has limited fidelity.

Our research was drawn to a single polarizing qubit tomography. However, this approach could be directly adopted to any other physical platform and to the reconstruction of the states of higher dimensions.

\section*{Acknowledgment}

The investigation was supported by Program no. 0066-2019-0005 of the Ministry of Science and Higher Education of Russia for Valiev Institute of Physics and Technology of RAS and partially supported by Russian Foundation of Basic Research, grant no. 18-37-00204.

\section*{References}
\bibliography{article}

\providecommand{\newblock}{}
\begin{thebibliography}{10}
\expandafter\ifx\csname url\endcsname\relax
  \def\url#1{{\tt #1}}\fi
\expandafter\ifx\csname urlprefix\endcsname\relax\def\urlprefix{URL }\fi
\providecommand{\eprint}[2][]{\url{#2}}

\bibitem{goldstein2017}
Goldstein D~H 2017 {\em Polarized Light\/} 3rd ed (CRC Press)

\bibitem{collet2005}
Collett E 2005 {\em Field Guide to Polarization\/} (SPIE)

\bibitem{slussarenko2019}
Slussarenko S and Pryde G~J 2019 {\em Appl. Phys. Rev.\/} {\bf 6} 041303

\bibitem{liao2017}
Liao S~K, Yong H~L, Liu C, Shentu G~L, Li D~D, Lin J, Dai H, Zhao S~Q, Li B,
  Guan J~Y {\em et~al.\/} 2017 {\em Nat. Photonics.\/} {\bf 11} 509

\bibitem{nielsen2000}
Nielsen M~A and Chuang I~L 2000 {\em Quantum Computation and Quantum
  Information\/} (Cambridge University Press)

\bibitem{bohr1996}
Bohr N 1996 Discussion with einstein on epistemological problems in atomic
  physics {\em Niels Bohr Collected Works\/} vol~7 (Elsevier) pp 339--381

\bibitem{lvovsky2009}
Lvovsky A~I and Raymer M~G 2009 {\em Reviews of Modern Physics\/} {\bf 81} 299

\bibitem{paris2004}
Paris M and Rehacek J 2004 {\em Quantum state estimation\/} vol 649 (Springer
  Science \& Business Media)

\bibitem{dariano2003}
D'Ariano G~M, Paris M~G and Sacchi M~F 2003 {\em Advances in Imaging and
  Electron Physics\/} {\bf 128} 206--309

\bibitem{banaszek2013}
Banaszek K, Cramer M and Gross D 2013 {\em New Journal of Physics\/} {\bf 15}
  125020

\bibitem{bogdanov2009}
Bogdanov {\relax Yu}~I 2009 {\em Journal of Experimental and Theoretical
  Physics\/} {\bf 108} 928--935

\bibitem{bogdanov2010}
Bogdanov {\relax Yu}~I, Brida G, Genovese M, Kulik S, Moreva E and Shurupov A
  2010 {\em Physical review letters\/} {\bf 105} 010404

\bibitem{bogdanov2011}
Bogdanov {\relax Yu}~I, Brida G, Bukeev I, Genovese M, Kravtsov K, Kulik S,
  Moreva E, Soloviev A and Shurupov A 2011 {\em Physical Review A\/} {\bf 84}
  042108

\bibitem{ghosh1999}
Ghosh G 1999 {\em Optics communications\/} {\bf 163} 95--102

\bibitem{bogdanov2013}
Bogdanov {\relax Yu}~I, Kalinkin A~A, Kulik S~P, Moreva E~V and Shershulin V~A
  2013 {\em New Journal of Physics\/} {\bf 15} 035012

\bibitem{bogdanov2014}
Bogdanov {\relax Yu}~I, Bantysh B~I, Kalinkin A~A, Kulik S~P, Moreva E~V and
  Shershulin V~A 2014 {\em Journal of Experimental and Theoretical Physics\/}
  {\bf 118} 845--855

\bibitem{williams1999}
Williams P~A 1999 {\em Applied Optics\/} {\bf 38} 6508--6515

\bibitem{hou2016}
Hou Z, Zhu H, Xiang G~Y, Li C~F and Guo G~C 2016 {\em JOSA B\/} {\bf 33}
  1256--1265

\bibitem{vallone2008}
Vallone G, Pomarico E, De~Martini F and Mataloni P 2008 {\em Laser Physics
  Letters\/} {\bf 5} 398--403

\bibitem{bogdanov2016}
Bogdanov {\relax Yu}~I, Bantysh B~I, Bogdanova N~A, Kvasnyy A~B and Lukichev
  V~F 2016 Quantum states tomography with noisy measurement channels {\em
  International Conference on Micro-and Nano-Electronics 2016\/} vol 10224
  (International Society for Optics and Photonics) p 102242O

\bibitem{bogdanov2018}
Bogdanov {\relax Yu}~I, Bantysh B~I, Bogdanova N~A and Lukichev V~F 2018 {\em
  Laser Physics\/} {\bf 28} 025204

\bibitem{bantysh2019}
Bantysh B~I, Fastovets D~V and Bogdanov {\relax Yu}~I 2019 High-fidelity
  quantum tomography with imperfect measurements {\em International Conference
  on Micro-and Nano-Electronics 2018\/} vol 11022 (International Society for
  Optics and Photonics) p 110222N

\bibitem{matlablib}
Quantum tomography by root approach
  \url{https://github.com/PQCLab/RootTomography} accessed: 2019-12-21

\bibitem{vrehavcek2004}
{\v{R}}eh{\'a}{\v{c}}ek J, Englert B~G and Kaszlikowski D 2004 {\em Physical
  Review A\/} {\bf 70} 052321

\bibitem{ma2016}
Ma X, Jackson T, Zhou H, Chen J, Lu D, Mazurek M~D, Fisher K~A, Peng X, Kribs
  D, Resch K~J {\em et~al.\/} 2016 {\em arXiv preprint arXiv:1601.05379\/}

\bibitem{mochon2006}
Mochon C 2006 {\em Physical Review A\/} {\bf 73} 032328

\bibitem{de2008}
de~Burgh M~D, Langford N~K, Doherty A~C and Gilchrist A 2008 {\em Physical
  Review A\/} {\bf 78} 052122

\bibitem{pancharatnam1956}
Pancharatnam S 1956 Generalized theory of interference and its applications.
  part i. coherent pencils {\em Proceedings of the Indian Academy of
  Sciences-Section A\/} vol~44 (Springer) pp 247--262

\bibitem{berry1984}
Berry M~V 1984 {\em Proceedings of the Royal Society of London. A. Mathematical
  and Physical Sciences\/} {\bf 392} 45--57

\bibitem{borovkov1999}
Borovkov A~A 1999 {\em Mathematical Statistics\/} (New York: CRC Press) ISBN
  978-9056990183

\bibitem{bogdanov2013lpl}
Bogdanov {\relax Yu}~I and Kulik S~P 2013 {\em Laser Physics Letters\/} {\bf
  10} 125202

\end{thebibliography}

\appendix

\section{Polarizing states tomography protocols}\label{app:protocols}

A quantum tomography protocol is usually described by a set of detecting observables or measurement operators \cite{bogdanov2010,bogdanov2011,vrehavcek2004,ma2016,mochon2006,de2008}. To define a protocol of polarizing state tomography one could list the rotation angles $\alpha$ and $\beta$ of HWP and QWP (see \Fref{fig:scheme}).

HWP and QWP perform the qubit rotation on the Bloch sphere by angles $\pi$ and $\pi/2$ respectively ($\alpha$ and $\beta$ define rotation axes). Corresponding unitary matrices have the following form:
\begin{equation*}
  HWP = \left(\begin{array}{cc}
    \cos2\alpha & \sin2\alpha \\
    \sin2\alpha & -\cos2\alpha
  \end{array}\right),
\end{equation*}
\begin{equation*}
  QWP = \frac{1}{\sqrt2}\left(\begin{array}{cc}
    \cos2\beta+i & \sin2\beta \\
    \sin2\beta & -\cos2\beta+i
  \end{array}\right).
\end{equation*}
Here $\alpha$ and $\beta$ are the fast axes angles with respect to the vertical axis. In these equations we have omitted the phase multipliers $i$ for simplicity. Nevertheless, we will return to the global phase below.

We assume that the photon passes the HWP first and then passes QWP. The total unitary matrix of polarization transformation is
\begin{equation*}
  U = QWP \cdot HWP = \frac{1}{\sqrt2}\left(\begin{array}{cc}
    \cos\gamma+i\cos\delta & -\sin\gamma+i\sin\delta \\
    \sin\gamma+i\sin\delta & \cos\gamma-i\cos\delta
  \end{array}\right).
\end{equation*}
Here we have introduced the following notation: $\gamma = 2(\beta-\alpha)$, $\delta = 2\alpha$.

Having a mixed quantum state $\rho$ in the input one would get the following probabilities of detectors $D_V$ and $D_H$ to produce a count (\Fref{fig:scheme}):
\begin{equation*}
  p_V = \braket{V| U \rho U^\dagger |V}, \quad p_H = \braket{H| U \rho U^\dagger |H}.
\end{equation*}.
It corresponds to the projective measurement of mutually orthogonal projectors $\Pi_V=\ketbra{\psi_V}$ and $\Pi_H=\ketbra{\psi_H}$ where
\begin{equation*}
  \ket{\psi_V} = U^\dagger\ket{V} = \frac{1}{\sqrt2}\left(\begin{array}{c} \cos\gamma-i\cos\delta \\ -\sin\gamma-i\sin\delta \end{array}\right),
\end{equation*}
\begin{equation*}
  \ket{\psi_H} = U^\dagger\ket{H} = \frac{1}{\sqrt2}\left(\begin{array}{c} \sin\gamma-i\sin\delta \\ \cos\gamma+i\cos\delta \end{array}\right).
\end{equation*}
At the same time, an arbitrary pure quantum state could be parameterized by spherical angles $\theta$ and $\varphi$ on the Bloch sphere:
\begin{equation}\label{eq:bloch_state}
  \ket{\psi} = \exp(i\chi)\left(\begin{array}{c} \cos(\theta/2) \\ \sin(\theta/2)\exp(i\varphi) \end{array}\right).
\end{equation}
Along with the spherical angles here we introduce the global phase $\chi$, which is, to some extent, analogous to geometrical (topological) Pancharatnam---Berry phase \cite{pancharatnam1956,berry1984}. The task of finding the correct combination of wave plates rotation angles $\alpha$ and $\beta$ by the spherical angles $\theta$ and $\varphi$ turns to the task of finding such a global phase $\chi$ that could equate $\ket{\psi}$ with $\ket{\psi_V}$. Let us note that one could represent the state $\ket{\psi_V}$ as the complex superposition of two vector states with real amplitudes:
\begin{equation*}
  \ket{\psi_V} = \frac{1}{\sqrt2}(\ket{\psi_\gamma}-i\ket{\psi_\delta}), \quad
  \ket{\psi_\gamma} = \left(\begin{array}{c} \cos\gamma \\ -\sin\gamma \end{array}\right), \quad
  \ket{\psi_\delta} = \left(\begin{array}{c} \cos\delta \\ \sin\delta \end{array}\right).
\end{equation*}
These vector states have unit normalization:
\begin{equation}\label{eq:state_norm}
  \braket{\psi_\gamma|\psi_\gamma} = \braket{\psi_\delta|\psi_\delta} = 1
\end{equation}
Let us write down \eref{eq:bloch_state} in the form
\begin{equation*}
  \ket{\psi} = \left(\begin{array}{c} \cos(\theta/2)\cos(\chi)+i\cos(\theta/2)\sin(\chi) \\ \sin(\theta/2)\cos(\varphi+\chi)+i\sin(\theta/2)\sin(\varphi+\chi) \end{array}\right).
\end{equation*}
It follows from the normalization condition \eref{eq:state_norm} that
\begin{equation*}
  \cos^2(\theta/2)\cos^2(\chi)+\sin^2(\theta/2)\cos^2(\varphi+\chi)=1/2,
\end{equation*}
\begin{equation*}
  \cos^2(\theta/2)\sin^2(\chi)+\sin^2(\theta/2)\sin^2(\varphi+\chi)=1/2
\end{equation*}
and
\begin{equation*}
  \tan2\chi = \frac{\cos^2(\theta/2)+\sin^2(\theta/2)\cos(2\varphi)}{\sin^2(\theta/2)\sin(2\varphi)}.
\end{equation*}
It gives
\begin{equation}\label{eq:glob_phase}
  \chi = \frac12 \arctan{\frac{\cos^2(\theta/2)+\sin^2(\theta/2)\cos(2\varphi)}{\sin^2(\theta/2)\sin(2\varphi)}} + \frac\pi2 k, \quad k=0,1,2,3.
\end{equation}
Equation \eref{eq:glob_phase} describes four possible solutions with $k = 0,1,2,3$. Any other $k$ does not give a new physical solution as it just shifts the global phase by an integer number of periods $2\pi$ in relation to previous solutions.

Next, we determine $\gamma$ from
\begin{equation*}
  \cos\gamma = \sqrt2\cos(\theta/2)\cos(\chi), \quad \sin\gamma = -\sqrt2\sin(\theta/2)\cos(\varphi+\chi).
\end{equation*}
We get $\gamma = \arccos(\cos\gamma)$ when $\sin\gamma \geq 0$ and $\gamma = -\arccos(\cos\gamma)$ otherwise. Similarly, for $\delta$ we have
\begin{equation*}
  \cos\delta = -\sqrt2\cos(\theta/2)\sin(\chi), \quad \sin\delta = -\sqrt2\sin(\theta/2)\sin(\varphi+\chi).
\end{equation*}
We get $\delta = \arccos(\cos\delta)$ when $\sin\delta \geq 0$ and $\delta = -\arccos(\cos\delta)$ otherwise. After derivation of $\gamma$ and $\delta$ we get desired angles: $\alpha=\delta/2$, $\beta=(\gamma+\delta)/2$. For unambiguity we consider these angles modulo $\pi$.

Let us consider an example of the cube symmetry measurement protocol, which has three different wave plates configuration. For $k=0$ one obtains
\begin{equation*}
\begin{array}{llll}
  \theta_1=\pi/2, & \varphi_1=0,      & \alpha_1=5\pi/8,     & \beta_1=\pi/2, \\
  \theta_2=\pi/2, & \varphi_2=\pi/2, & \alpha_2=11\pi/16, & \beta_2=3\pi/4, \\
  \theta_3=0,      & \varphi_3=0,      & \alpha_3=\pi/2,      & \beta_3=\pi/2. \\
\end{array}
\end{equation*}
Let us also consider the octahedron symmetry measurement protocol. The corresponding projection states are located at the centers of the octahedron circumscribed about the Bloch sphere. We derive four wave plates configurations for the states $\ket{\psi_V}$ located at the top semi-sphere of the Bloch sphere ($k=0$):
\begin{equation*}
\begin{array}{llll}
  \theta_1=\arccos1/\sqrt3, & \varphi_1=\pi/4,   & \alpha_1\approx1.9210, & \beta_1\approx1.8785, \\
  \theta_2=\arccos1/\sqrt3, & \varphi_2=3\pi/2, & \alpha_2\approx2.7914, & \beta_2\approx2.8339, \\
  \theta_3=\arccos1/\sqrt3, & \varphi_3=5\pi/4, & \alpha_3\approx1.2206, & \beta_3\approx1.2631, \\
  \theta_4=\arccos1/\sqrt3, & \varphi_4=7\pi/4, & \alpha_4\approx0.3502, & \beta_4\approx0.3077. \\
\end{array}
\end{equation*}
The rest four points of the octahedron edges correspond to the states $\ket{\psi_H}$.

For the 2-qubit state tomography one could place a pair of wave plates (HWP and QWP) at each of the two radiation spatial modes. Then the measurement protocol is defined by all the combinations of wave plates configurations in these modes. The 2-qubit cube symmetry protocol has $3\cdot3=9$ configurations in total ($9\cdot4=36$ measurements operators, that correspond to the different projections of the quantum state). Similarly, the 2-qubit octahedron symmetry protocol has $4\cdot4$ configurations ($16\cdot4=64$ operators).

\section{Simulation results for the octahedron symmetry protocol}\label{app:octahedron}

Figures below demonstrate the results of the simulation of polarizing qubit quantum tomography in presence of chromatic aberration. The results were obtained for the octahedron symmetry measurement protocol (\ref{app:protocols}).

\begin{figure}[h]
  \centering
  \includegraphics[width=.7\linewidth]{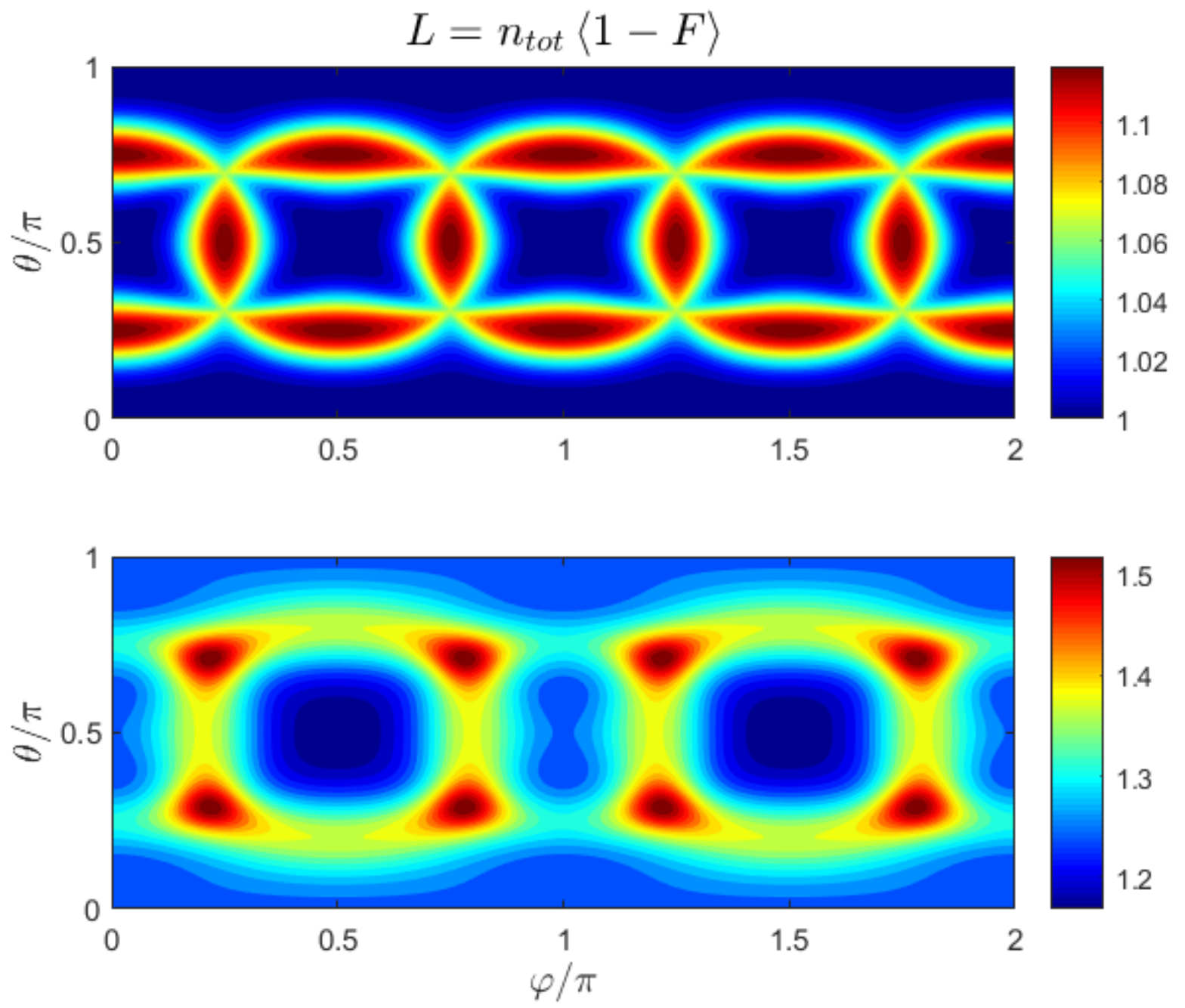}
  \caption{Loss function distribution over pure single-qubit states $\cos(\theta/2)\ket{0}+\sin(\theta/2)e^{i\varphi}\ket{1}$. The octahedron symmetry protocol. Top --- the case of monochromatic radiation ($\Delta \lambda = 0$), bottom --- $\Delta \lambda = 0.01$\textmu m.}
  \label{fig:LBloch_octa}
\end{figure}

For $\Delta\lambda= 0.01$\textmu m (\Fref{fig:LBloch_octa}) we obtained $L_{min} = 1.1679$, $L_{max} = 1.5360$. Here the minimal fidelity loss (1.1679) is higher than the maximal one in the case of monochromatic radiation (1.125), just as for the case of the cube symmetry protocol. The whole distribution is also rearranged.

\begin{figure}[h]
  \centering
  \begin{minipage}{0.48\textwidth}
    \centering
    \includegraphics[width=\textwidth]{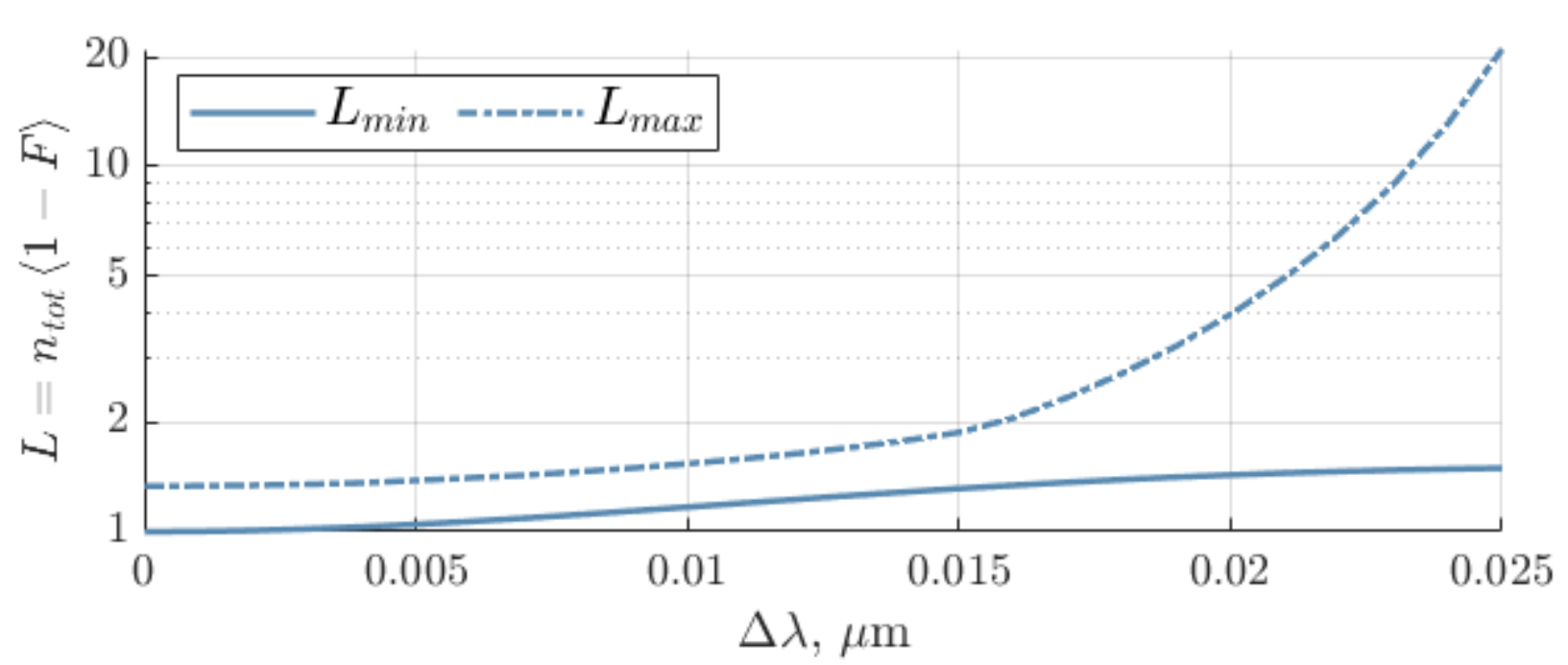}\\(a)
  \end{minipage}\hfill
  \begin{minipage}{0.48\textwidth}
    \centering
    \includegraphics[width=\textwidth]{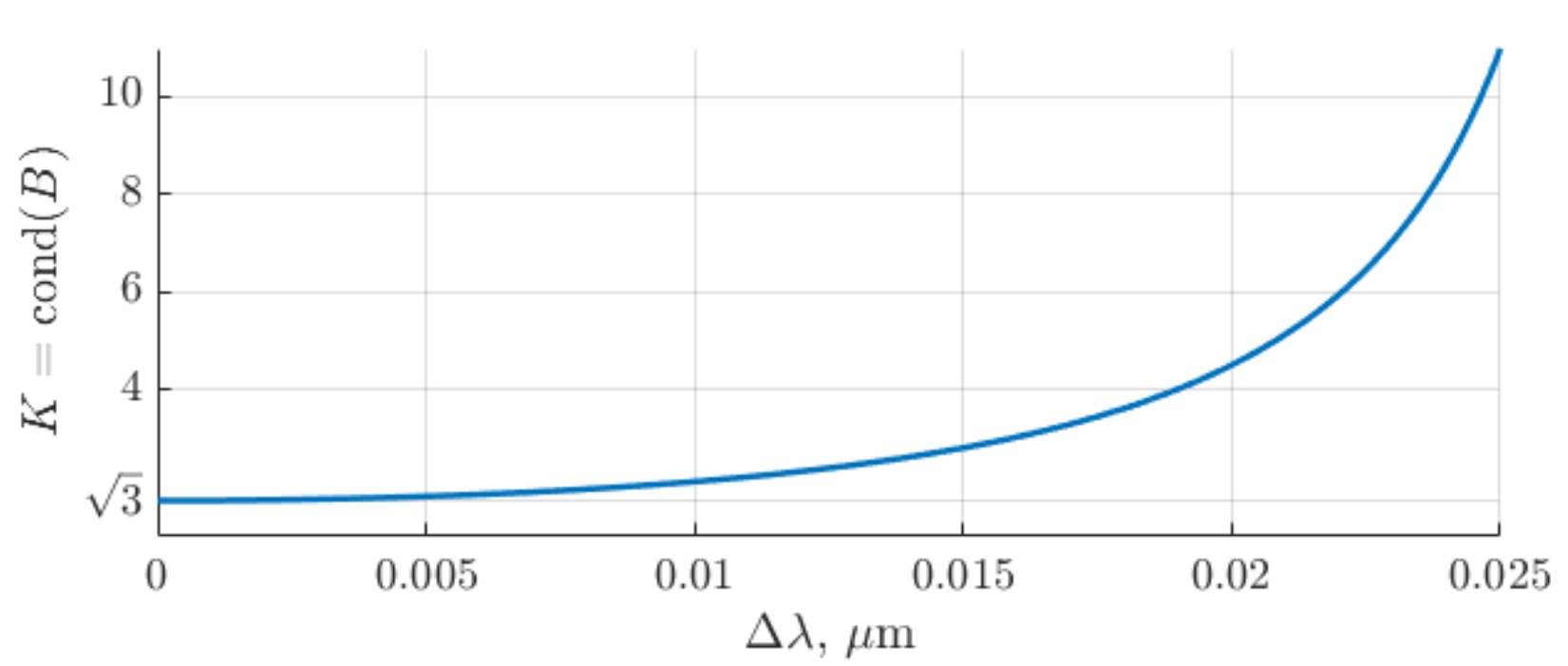}\\(b)
  \end{minipage}
  \caption{Minimal and maximal fidelity loss function (a) and measurement matrix condition number (b) vs. the radiation spectral bandwidth. The octahedron symmetry protocol.}
  \label{fig:Lminmax_cond_octa}
\end{figure}

In \Fref{fig:Lminmax_cond_octa} we show the dependencies of minimal and maximal values of the loss functions and measurement matrix condition number on the radiation spectral bandwidth. In the limit $\Delta\lambda \rightarrow 0$ maximal fidelity loss approaches $4/3$. This value is higher than the maximal value 1.125 for the model of projective measurements when $\Delta\lambda = 0$. This deviation is observed only within infinitesimal neighborhoods of eight points corresponding to the ideal projectors of octahedron symmetry protocol.

\begin{figure}[h]
  \centering
  \begin{minipage}{0.48\textwidth}
    \centering
    \includegraphics[width=\textwidth]{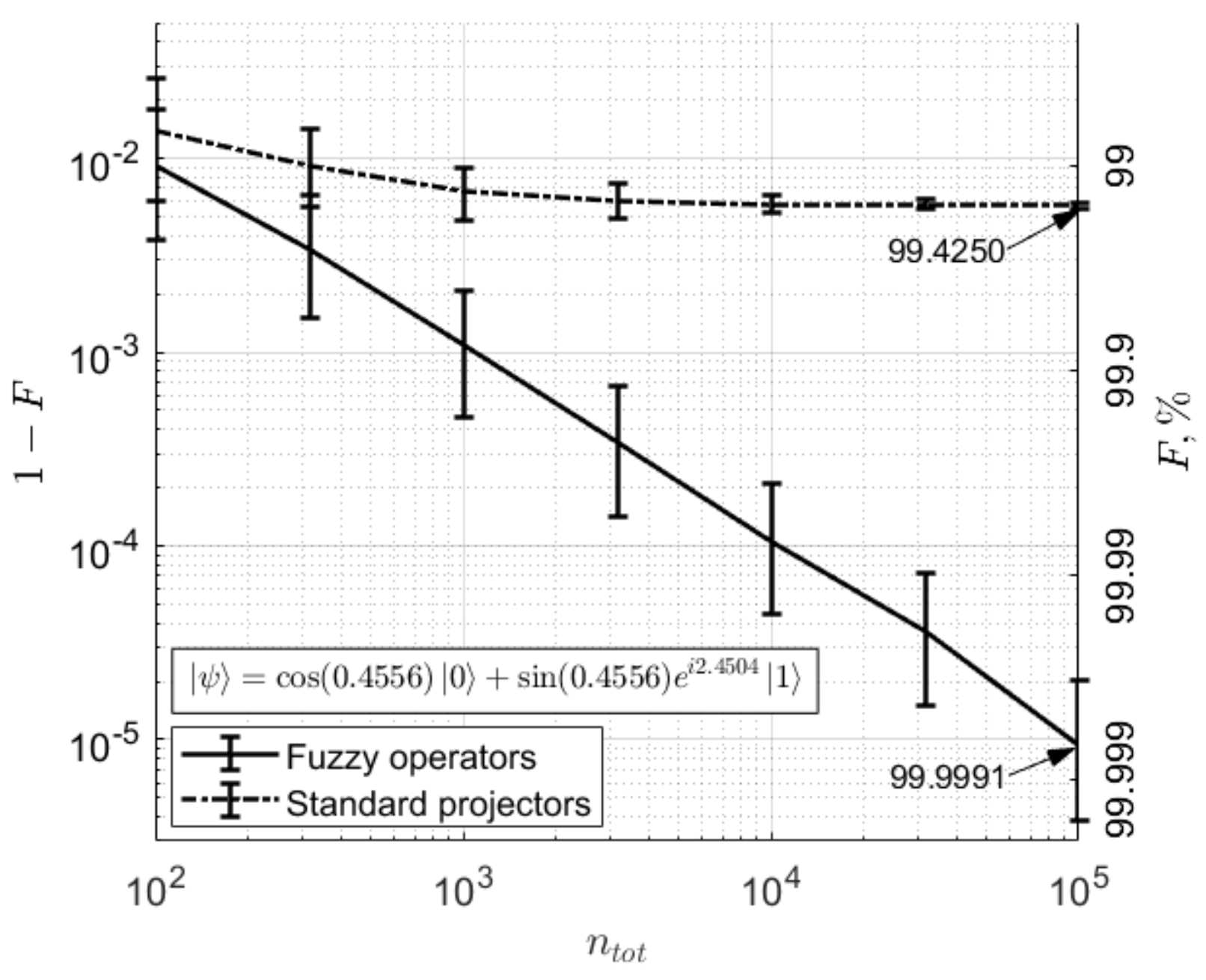}\\(a)
  \end{minipage}\hfill
  \begin{minipage}{0.48\textwidth}
    \centering
    \includegraphics[width=\textwidth]{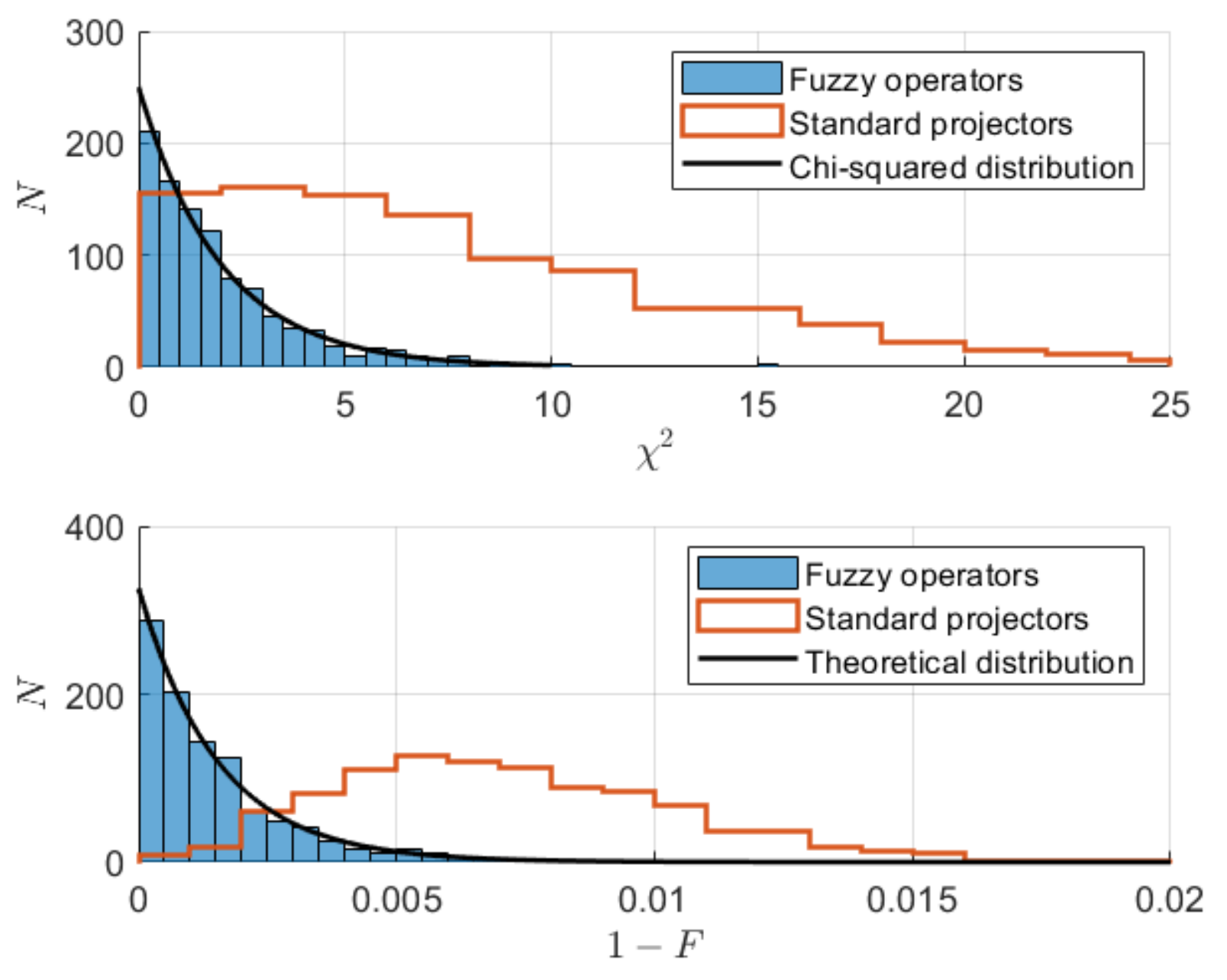}\\(b)
  \end{minipage}
  \caption{The results of Monte Carlo simulation of the quantum state tomography in presence of chromatic aberration. The octahedron symmetry protocol. Radiation spectral bandwidth $\Delta\lambda = 0.01$\textmu m. Comparison of fuzzy measurements model at $\Delta\lambda = 0.01$\textmu m and standard model of projective measurements. (a) The reconstruction fidelity vs. total sample size. The plot shows median and lower and upper quartile over 1000 numerical experiments. (b) The result of 1000 numerical experiments with $n_{tot}=1000$ in each. Top --- histograms of chi-squared for two models and theoretical chi--squared distribution for 1 degree of freedom. Bottom --- histograms of infidelity for two models and theoretical distribution based on information matrix.}
  \label{fig:fid_octa}
\end{figure}

The results in \Fref{fig:fid_octa} were obtained for the state with $\theta=0.9111$, $\varphi=2.4504$. This state is characterized by the value of loss function close to the maximal one (according to the results in \Fref{fig:LBloch_octa}).

Again, one could observe that the fuzzy measurement model gives the reconstruction fidelity close to the theoretical limit calculated via information matrix. At the same time, the use of standard model of projective measurements results in much higher values of fidelity losses.

\section{Root approach to quantum tomography}\label{app:root}

\subsection{Quantum state reconstruction}

Consider a mixed state in the Hilbert space of dimension $s$. The state density matrix $\rho$ could be represented as a pure vector state in the space of a higher dimension \cite{nielsen2000}. We denote its complex amplitudes as a column vector $c$. One could get $c$ from $\rho$ in the following way. Let us construct a block matrix $\psi = (\sqrt{\lambda_1}v_1, \dots, \sqrt{\lambda_r}v_r)$ of dimension $s \times r$, where $r$ is the rank of the density matrix, $\lambda_i$ --- its non-zero eigenvalues, $v_i$ --- corresponding eigenvectors. Column vector $c$ of the purified state is obtained by reshaping rectangular matrix $\psi$ into a single column (the second column is put under the first one and so on). Note that the state descriptions using $\psi$ and $c$ are completely equavalent. It only differs in the way of listing the complex amplitudes.

It is obvious that $\rho=\psi\psi^\dagger$, hence the purification is ambiguous and is defined up to a unitary gauge. In particular, matrices $\psi$ and $\psi V$ ($V$ is a unitary matrix of dimension $r \times r$) describe the same quantum state as they give the same density matrix. In the case of a pure state ($r=1$) this ambiguity is related to the gauge invariance of a vector state global phase.

The use of purified matrix $\psi$ has advantage in quantum state reconstruction by maximum likelihood estimation as the resulting density matrix is always non-negative \cite{banaszek2013,bogdanov2009,bogdanov2010}. The statistical data processing implies numerical solving of the quasilinear likelihood equation:
\begin{equation}\label{eq:maxlik_eq}
  I\psi = J(\psi)\psi, \quad \textnormal{where} \quad I=\sum_j{n_j \Lambda_j}, \quad J(\psi)=\sum_j{\frac{k_j}{p_j(\psi)}\Lambda_j}.
\end{equation}
Here $k_j$ is the number of observed experimental counts corresponding to the measurement operator $\Lambda_j$, $p_j(\psi)=\Tr(\psi\psi^\dagger\Lambda_j)$ --- probability to observe the corresponding event, $n_j$ --- number of corresponding measurement trials. The summations in \eref{eq:maxlik_eq} are performed over all $l \cdot s$ measurement operators of $l$ independent measurement schemes. In the current work we consider POVM-measurements, which give $I = n_{tot}E$, where $n_{tot}$ is the total sample size and $E$ is the identity matrix of dimension $s \times s$.

Let us note that the maximum likelihood estimator is asymptotically efficient \cite{borovkov1999}. Hence, one could obtain quantitative theoretical estimations of the quantum state reconstruction fidelity using information matrix (\ref{app:fidelity}).

\subsection{Reconstruction adequacy}

Let $\psi$ (and corresponding density matrix $\rho=\psi\psi^\dagger$) be the result of a single tomography experiment. One could perform the adequacy estimation of the model using chi-squared criterion \cite{bogdanov2009,bogdanov2013lpl}. Let us calculate the variable
\begin{equation*}
  \chi^2 = \sum_j{\frac{(k_j-n_jp_j(\psi))^2}{n_jp_j(\psi)}}.
\end{equation*}
Here $n_jp_j$ ($k_j$) are the expected (observed) counts.

According to the general theory, for asymptotically optimal statistical estimators $\chi^2$ is the chi-squared distributed random variable \cite{borovkov1999}. The number of degrees of freedom $\nu$ is determined by $\nu = ls-\nu_P-\nu_{norm}$. Here $\nu_P=(2s-r)r-1$ is the number of independent real parameters characterizing the quantum state ($s$ --- Hilbert space dimension, $r$ --- rank of the quantum state density matrix); $\nu_{norm}$ --- number of normalization conditions: $\nu_{norm}=1$ if all the measurement operators form a single POVM--measurement, $\nu_{norm}=l$ for the case of $l$ independent POVM--measurements.

Maximum likelihood estimator is among the asymptotically optimal estimators. In the language of statistical information the asymptotic condition is $h_k \gg 1$ for all $k = 1,\dots,\nu_P$. Here $h_k$ are the eigenvalues of information matrix $H$ (see \ref{app:fidelity}).

The p--value of the chi-squared test is determined by the area under the curve of the chi-squared distribution probability density function to the right of the obtained $\chi^2$:
\begin{equation*}
  \textnormal{p--value} = \textnormal{Pr} \lshad X \geq \chi^2 | \nu \rshad, 
\end{equation*}
where $X$ is the chi-squared random variable with $\nu$ degrees of freedom.

\section{Informational approach to the quantum tomography fidelity analysis}\label{app:fidelity}

Quantum state tomography fidelity depends on the state under research and the measurement protocol. One could obtain detailed fidelity characteristics before performing any measurements by using analytical information theory \cite{bogdanov2009,bogdanov2010}. The notations below are defined in \ref{app:root}

Let us consider a purified vector state in a Hilbert space of dimension $rs$. Its complex amplitudes form a column vector $c$. Let us move to a real Euclidean space of dimension $2rs$:
\begin{equation*}
  c \rightarrow \left(\begin{array}{c} \Re(c) \\ \Im(c) \end{array}\right).
\end{equation*}
The measurement operators should be transformed in the following way:
\begin{equation*}
  \Lambda_j
  \rightarrow
  \underbrace{\left(\begin{array}{ccc} \Lambda_j & 0 & 0 \\ 0 & \ddots & 0 \\ 0 & 0 & \Lambda_j \end{array}\right)}_\textnormal{\footnotesize $r$ blocks}
  \rightarrow
  \left(\begin{array}{cc} \Re(\Lambda_j) & -\Im(\Lambda_j) \\ \Im(\Lambda_j) & \Re(\Lambda_j) \end{array}\right).
\end{equation*}

The fundamental probabilistic nature of quantum measurements leads to statistical fluctuations of reconstructed quantum state parameters. The level of these fluctuation could be described in terms of complete information matrix \cite{bogdanov2009,bogdanov2010,bogdanov2011}:
\begin{equation*}
  H = 2\sum_j{n_j\frac{(\Lambda_j c)(\Lambda_j c)^T}{p_j}}.
\end{equation*}
This matrix is symmetric, non-negative and has dimension $2rs \times 2rs$. In the case of complete tomography protocols that form the unity decomposition ($\sum_j{n_j\Lambda_j} \propto E$) $H$ has one eigenvalue equal to $2n_{tot}$ and $r^2$ zero eigenvalues. The rest of $\nu_P=(2s-r)r-1$ eigenvalues determine the amount of information about $c$, which could be extracted from the measurements. Let us denote these eigenvalues as $h_k$ ($k=1,\dots,\nu_P$) and define
\begin{equation*}
  d_k = \frac{1}{2h_k}, \quad k=1,\dots,\nu_P.
\end{equation*}
The infidelity of the purified state reconstruction is the random variable that has generalized chi-squared distribution:
\begin{equation*}
  1-F = \sum_{k=1}^{\nu_P}{d_k \xi_k^2},
\end{equation*}
where each random variable $\xi_k$ has the standard normal distribution. The mean value of infidelity is $\braket{1-F}=\sum_k{d_k}$, the variance is $\sigma_{1-F}^2=2\sum_k{d_k^2}$. Let us note that the information matrix is proportional to the total sample size $n_{tot}$. Hence, $\braket{1-F}$ is inversely proportional to $n_{tot}$, so it is convenient to define the loss function as $L=n_{tot}\braket{1-F}$. It does not depend on the sample size but on the quantum state under research and the efficiency of the measurement protocol.

This asymptotic fidelity theory becomes adequate when the amount of information about every degree of freedom of the quantum state is high enough: $h_k \gg 1$ for every $k=1,\dots,\nu_P$. In this case $d_k \ll 1$ and $\braket{1-F} \ll 1$.

\end{document}